# THE SIMPLE NONPOLAR CONTINUUM MEDIA.
# PART I. THE EQUIVALENCE TRANSFORMATION.

V.O. Bytev

*10 Semakova Street, Tyumen State University, Tyumen, Russia, 625003, e-mail: vbytev@utmn.ru*
*Head of Appl. Math. Dep. Tyumen State University*

In the realm of Continuum Physics, material bodies are realized as continous media and so-called "extensive quantities", such as mass, momentum and energy, are monitored through the fields of their densities, which are related by balance laws constitutive equations.

While building the mathematical models of any continuum media mechanics, as a rule one uses three levels of description: phenomenological, dynamic, and statistical. We have deal with the phenomenological descriptive level.

Let us set the following equations as the system $S$:

$$\rho_t + div(\rho \vec{u}) = 0 \qquad (1)$$

$$\rho[\vec{u}_t + (\vec{u} \cdot \nabla)\vec{u}] - div\,\Pi(\nabla \vec{u}) + \nabla p = 0 \qquad (2)$$

$$\rho_t + (\vec{u} \cdot \nabla) + G(p,\rho)\,div\,\vec{u} + H(p,\rho)\Phi = 0 \qquad (3)$$

Here $\rho$ is density of the continuum medium considered; $p$ is the hydrostatic (equilibrium) pressure; $\vec{u}$ is the velocity vector; $\Pi$ is the tensor of viscous tension; $G(p,\rho) = -\rho S_\rho / S_p$; $H(p,\rho) = -\dfrac{1}{\rho \Theta S_p}$; $S = S(p,\rho)$ is the entropy; $\Theta = \Theta(p,\rho)$ - absolute temperature; and $\Phi = (\Pi : \vec{u})$ is the dissipative function. Allow $\Pi$ to be a symmetric tensor, depending only on $grad\,\vec{u}$. Such class of continuum media was named by C. Truesdell as simple nonpolar continuum media [1].

It is obviously the system of differential equations (1)-(3) is not closed, because the functions $\Pi^{ij}$, $G$, and $H$ are arbitrary.

So, we should try to find out some quite general approaches to the problem of determining equations necessary for the closure of a differential equations system the consequences of conservations lows. Apparently, it should be the principle of invariance. Further, it seems quite naturally to apply a group of continuum transformations in order to formulate the invariance principle itself.

Thus, we have the problem of group classification of differential consequences of conservation lows. The first problem in this way is the problem of finding equivalence transformations [2,3]. This problem consists of the constructions of the space $R^{n+m+A}(t,x,u,p,\rho,\Pi^{ij},\Pi^{ij}_{kl},G,H)$ that presents the system of differential equations (1)-(3), while only changing their representative $\Pi^{ij}$, $\Pi^{ij}_{kl}$, $G$, $H$.

Here $\Pi^{ij}$ are components of the tensor $\Pi$, and $\Pi^{ij}_{kl} = \dfrac{\partial \Pi^{ij}}{\partial p^k_l}$, $p^k_l = \dfrac{\partial u^k}{\partial x^l}$, $(i,j,k,l = 1,2,...,N; N = 1,2,3)$.

As usually, for this purpose a one-parameter Lie's group of transformations of the space $R^{n+m+A}$ with the group parameter $a$ is used. For more details one can see [3].

The generator of this group has the form

$$X = \xi^t \frac{\partial}{\partial t} + \xi^{x_i} \frac{\partial}{\partial x_i} + \eta^{u_i} \frac{\partial}{\partial u_i} + \eta^p \frac{\partial}{\partial p} + \eta^\rho \frac{\partial}{\partial \rho} + \mu^{kt} \frac{\partial}{\partial \Pi^{kt}} + \mu^{kt}_{es} \frac{\partial}{\partial \Pi^{kt}_{es}} + $$
$$+ \mu^G \frac{\partial}{\partial G} + \mu^H \frac{\partial}{\partial H} \qquad (4)$$

In the classical approach [2] for equivalence groups it was assumed $\xi^t = \xi^t(t,x,u,p,\rho)$, $\xi^{x_i} = \xi^{x_i}(t,x,u,p,\rho)$; $\eta^{u_i} = \eta^{u_i}(t,x,u,p,\rho)$, $\eta^\rho = \eta^\rho(t,x,u,p,\rho)$, $\eta^p = \eta^p(t,x,u,p,\rho)$.

The main result is Theorem. In classical approach the generators (4) are

$$X_0 = \frac{\partial}{\partial t}, \; X_i = \frac{\partial}{\partial x_i}, \; S = \frac{\partial}{\partial p}, \; Y_i = t\frac{\partial}{\partial x_i} + \frac{\partial}{\partial u_i}; \; T = \sum_{k=1}^{N} \frac{\partial}{\partial \Pi^{kk}} - H\frac{\partial}{\partial G};$$
$$Z_1 = x^i \frac{\partial}{\partial x^i} + u^j \frac{\partial}{\partial u^j} + 2p\frac{\partial}{\partial p} + 2\Pi^{kt}\frac{\partial}{\partial \Pi^{kt}} + 2\Pi^{kt}_{es}\frac{\partial}{\partial \Pi^{kt}_{es}} + 2G\frac{\partial}{\partial G};$$
$$Z_2 = \rho\frac{\partial}{\partial \rho} + p\frac{\partial}{\partial p} + \Pi^{kt}\frac{\partial}{\partial \Pi^{kt}} + \Pi^{kt}_{es}\frac{\partial}{\partial \Pi^{kt}_{es}} + G\frac{\partial}{\partial G}.$$

This means that the equivalence group of model (1)-(3) does not have $SO_3$ as a subgroup.

Henceforth, we should take it in account in the time of building any models of continuum media.

Moreover, we can not now assume isotropy of tensor $\Pi$.

Conversely, we should solve the problem of group classification for system (1)-(3).

It will be the part II.